\documentclass[twocolumn]{elsart}

\usepackage{natbib}

\usepackage{graphicx}

\usepackage{amssymb}

\begin{document}

\begin{frontmatter}


 \title{Impurity Scattering Effects in STM Studies of High-$T_{\rm c}$ Superconductors}
\author[label1]{Noboru Fukushima}
\ead{noboru@phys.sinica.edu.tw},
\author[label1,label2]{Chung-Pin Chou},
\author[label1]{Ting-Kuo Lee} 
\address[label1]{Institute of Physics, Academia Sinica, NanKang, Taipei 11529, Taiwan}
\address[label2]{Department of Physics, National Tsinghua University, Hsinchu 300, Taiwan}



%
%

\begin{abstract}
Recent STM measurements have observed many inhomogeneous patterns of the
local density of states on the surface of high-$T_{\rm c}$ cuprates.  As
a first step to study such disordered strong correlated systems, we use
the BdG equation for the $t$-$t'$-$t''$-$J$ model with an impurity.  The
impurity is taken into account by a local potential or local variation
of the hopping/exchange terms.
Strong correlation is treated by a Gutzwiller mean-field
theory with local Gutzwiller factors and local chemical potentials.
It turned out that the potential impurity scattering is greatly
suppressed, while the local variation of hoppings/exchanges is enhanced.
\end{abstract}


\end{frontmatter}

\section{Background}

Anderson's theorem tells us that the s-wave superconductivity
is insensitive to small potential scattering.
On the other hand, the d-wave superconductivity has
zero superconducting gap in the nodal direction, and thus
may be sensitive to disorder.
However, experimental observation of
the high-temperature superconductivity,
which many people are nowadays convinced has d-wave symmetry,
seems robust against disorder.
For example,
local density of states measured by the STM
\cite{Kohsaka07}
show clear V-shape at low energy
that indicates the d-wave nodes are not much influenced by disorder.
Theoretically,
it is proposed that this protection of V-shape is due to
strong Coulomb repulsion between electrons \cite{Anderson00}.
Hence detailed studies of effects of strong correlation for impurity scattering are
necessary, and in this paper we focus on it.

In the Gutzwiller approximation,
the model parameters are replaced by renormalized ones
in return for taking the intractable projection operator away.
Then, $t$-term is renormalized by a factor $g_t<1$
because hopping is more difficult in the presence of projection.
On the other hand, $J$-term is renormalized by $g_s > 1$ because each
site is more often singly occupied.
In this paper, we focus on another term;
how are impurity terms renormalized?

\section{Model}

We use $t$-$t'$-$t''$-$J$ model with an impurity term, namely,
\begin{equation}
H = H_t + H_J +H_{\rm imp},
\end{equation}
\begin{equation}
 H_{t} = P_{\rm G}\left(- \sum_{i,j,\sigma} t_{ij}  c_{i\sigma}^\dagger
		   c_{j\sigma} \right)P_{\rm G},
\end{equation}
\begin{equation}
 H_{J} = J \sum_{\langle i,j\rangle}\vec{S}_i\cdot \vec{S}_j,
\end{equation}
where $t_{ij}=t, t', t'',$ for nearest, second, third neighbors,
respectively, and otherwise zero. The summation in the $J$ term is taken
over every nearest-neighbor pair.  The Gutzwiller projection operator $P_{\rm G}$
prohibits electron double occupancy at every site.
Throughout this paper we take $t'=-0.3t, t''=0.2t, J=0.28t,$ and the hole
density $x=0.125$.

We put an impurity at $i=0$.
Here, we try three different types of the impurity term and compare them:
(i) impurity potential,
\begin{equation}
 H_{\rm imp} = P_{\rm G}\left(V_0\sum_\sigma c_{0\sigma}^\dagger c_{0\sigma} \right)P_{\rm G},
\end{equation}
(ii) local $t$ variation,
\begin{eqnarray}
&& H_{\rm imp} =  \nonumber \\
&& - P_{\rm G} \sum_{j\sigma} \delta t_{0j} \left( 
 c_{0\sigma}^\dagger c_{j\sigma} + c_{j\sigma}^\dagger c_{0\sigma}
\right)P_{\rm G},
\end{eqnarray}
(iii) local $J$ variation,
\begin{equation}
H_{\rm imp} =  \delta J \sum_{j{\rm (n.n.)}} \vec{S}_0\cdot \vec{S}_j.
\end{equation}

\section{Method}

We solve a Bogoljubov-de Gennes (BdG) equation using
the Gutzwilller approximation
with local Gutzwiller factors and local chemical potentials
\cite{QHWang06,CLi06}.
Let us assume that a good variational ground state can be
represented in the form of $P|\psi\rangle$,
where $|\psi\rangle$ represents a state obtained by solving
a BdG equation. The projection $P$ includes $P_G$ and an operator
to control the particle number.

The Gutzwiller approximation
assuming that the average of
the local electron density $\hat{n}_i=\sum_{\sigma} c_{i\sigma}^\dagger c_{i\sigma}$ 
is not changed by the projection, {\it i.e.},
\begin{equation}
\frac{\langle \psi|P \hat{n}_i P|\psi\rangle}
     {\langle \psi|PP|\psi\rangle}
=
\frac{\langle \psi| \hat{n}_i |\psi\rangle}
     {\langle \psi|\psi\rangle},
\label{eq:niconserve}
\end{equation}
gives the local Gutzwiller factors as
\begin{equation}
g^{t}_{ij}=\sqrt{\frac{2x_i}{1+x_i}} \sqrt{\frac{2x_j}{1+x_j}},
\label{eq:gutzt}
\end{equation}
\begin{equation}
g^{s}_{ij}=\frac{2}{1+x_i}\cdot\frac{2}{1+x_j},
\label{eq:gutzs}
\end{equation}
where the local hole density
$x_i=1-\langle \hat{n}_i \rangle_0$. Here,
$\langle \ldots \rangle_0$ denotes the expectation value by
$|\psi\rangle$.
From the extremum condition the free energy,
the following Bogoljubov-de Gennes (BdG) equation is derived:
\begin{eqnarray}
&& H_{\rm BdG} =
- \sum_{ij}  g^{t}_{ij} t_{ij} c_{i\sigma}^\dagger c_{j\sigma}\nonumber\\&&
- \sum_{\langle ij \rangle} \frac34 g^{s}_{ij} J \times \nonumber\\&&
\Big\{ \chi_{ij}(
c_{i\uparrow}^\dagger c_{j\uparrow}  +  c_{i\downarrow}^\dagger c_{j\downarrow}
+ {\rm H.c.} ) \nonumber\\&&
+ \Delta_{ij}( c_{i\uparrow}^\dagger c_{j \downarrow}^\dagger +
c_{j\uparrow}^\dagger c_{i \downarrow}^\dagger
+ {\rm H.c.} )
\Big\} \nonumber\\&&
-\sum_{i\sigma} (\mu+\mu_i)  c_{i\sigma}^\dagger c_{i\sigma} +H_{\rm imp},
\label{eq:BdG}
\end{eqnarray}
where $
\chi_{ij}
= \langle c_{i\uparrow}^\dagger c_{j\uparrow} \rangle_0 
= \langle c_{i\downarrow}^\dagger c_{j\downarrow} \rangle_0
$, $ \Delta_{ij}=\langle c_{j\downarrow} c_{i\uparrow} \rangle_0$.
We assume that $\chi_{ij}$ and $\Delta_{ij}$ are real numbers and
that $\Delta_{ij}=\Delta_{ji}$.
The local chemical potential $\mu_i$ is the derivative of the internal energy
with respect to the local hole density,
\begin{eqnarray}
 \mu_i &=& -\sum_{j} 4 \frac{dg^{t}_{ij}}{dx_i} t_{ij} \chi_{ij}
  \nonumber \\ &&
-\sum_{j \rm{(n.n.)}}\frac32 \frac{dg^{s}_{ij}}{dx_i}
J(\chi_{ij}^2 + \Delta_{ij}^2).
\end{eqnarray}
We use a supercell composed of 20$\times$20 sites whose origin is an impurity site.
This supercell is repeated to construct a superlattice of 10$\times$10
supercells with the periodic boundary condition \cite{Tsuchiura00}.
Then, the Hamiltonian can be block-diagonalized by the Fourier transform
with respect to the supercell indices,
and calculation of expectation values is reduced to
an average over many twisted boundary conditions of the
20$\times$20 site system.

\section{Results and Discussion}

Let us start from impurity type (i), namely, the impurity potential.
As mentioned in Eq.~(\ref{eq:niconserve}),
$n_i$ is not renormalized by any $g$ factor by definition.
However, as one can see in the BdG Hamiltonian, Eq.~(\ref{eq:BdG}),
the impurity potential can be compensated by $\mu_i$.
Therefore, we define a renormalized impurity potential by including
difference of $\mu_{i}$, namely, 
\begin{equation}
 \tilde{V_0}=V_0-\left(\mu_{i=0}-\mu_{\infty} \right)
\end{equation}
where $\mu_{\infty}$ is the local chemical potential at the site farthest from $i=0$.

Figure 1 shows $\tilde{V}_0$ as a function of $V_0$.
Note that $\tilde{V}_0$ is strongly suppressed, and
the renormalization factor seems about the order of $g^t=2x/(1+x)$
which is plotted by a dotted line for comparison.
In our understanding, it can be explained as follows:
Basically, impurity sites are uncomfortable for electrons to stay at.
However, under strong electron correlation, everywhere is uncomfortable
and impurity sites may be less uncomfortable compared to the systems
without correlation.  This effect appears as compensation of the
impurity potential by local chemical potentials.

\begin{figure}[h]
 \begin{center}
\includegraphics[width=7cm]{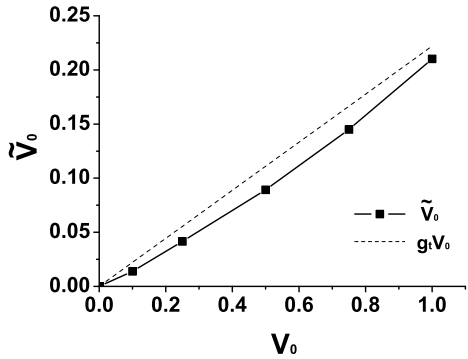}
 \end{center}
\caption{The renormalized impurity potential as a function of the bare
 impurity potential.}
\end{figure}

Next, in order to see the influence from the type-(ii) and (iii) impurity,
we plot the local electron density along $x$-axis in Fig.~2.
Here, Fig.~2(a) is for the type-(ii) impurity with
$\delta t/t = \pm 0.1$.
When $t$ is smaller (larger) locally, the electron density near the
impurity site is higher (lower).
Therefore, effective $t_{ij}$, namely, $g^t_{ij} t_{ij}$, is further smaller (larger).
This should be because the system try to gain energy 
by increasing local Gutzwiller factor at large $t$ region.
Namely, this defect is enhanced by the strong correlation.

We also find similar behavior for type-(iii) impurity as shown in Fig.~2(b),
where $\delta J/J = \pm 0.1$.
When $J$ is smaller (larger) locally, the electron density near the
impurity site is lower (higher), and
$g^s_{ij} J$, is further smaller (larger).
However, the magnitude of the enhancement is much smaller than that in
Fig.~2(a).  This should be because ${{\rm d}g^{s}_{ij}}/{{\rm d}x_i}$ is much
smaller than ${{\rm d}g^{t}_{ij}}/{{\rm d}x_i}$, and the effect by local electron
density modification is also much smaller in the $J$ variation than in
the $t$ variation.

\begin{figure}[h]
 \begin{center}
\includegraphics[width=7cm]{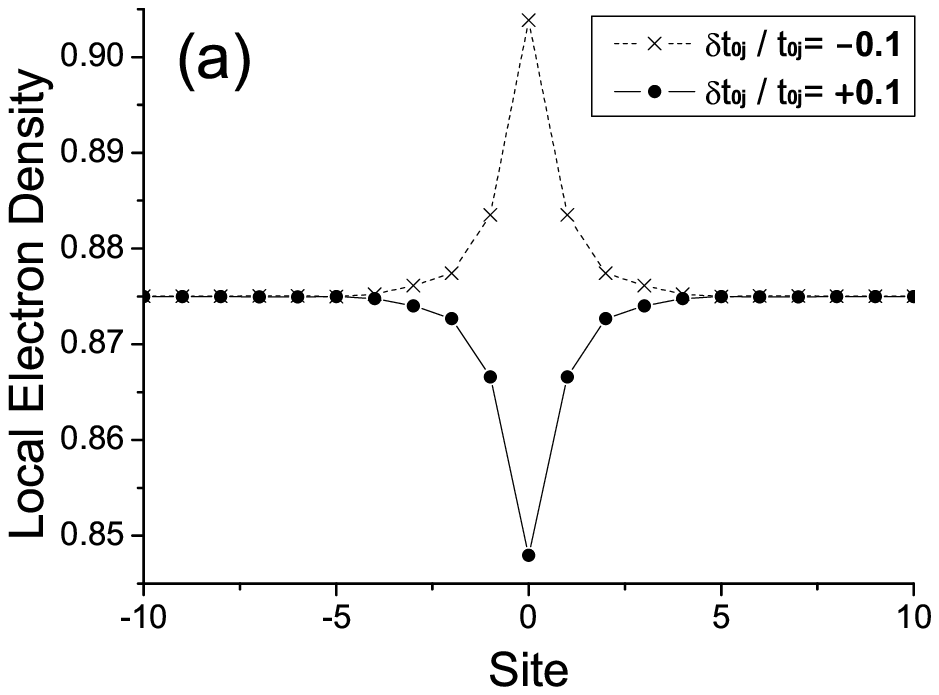}

\includegraphics[width=7cm]{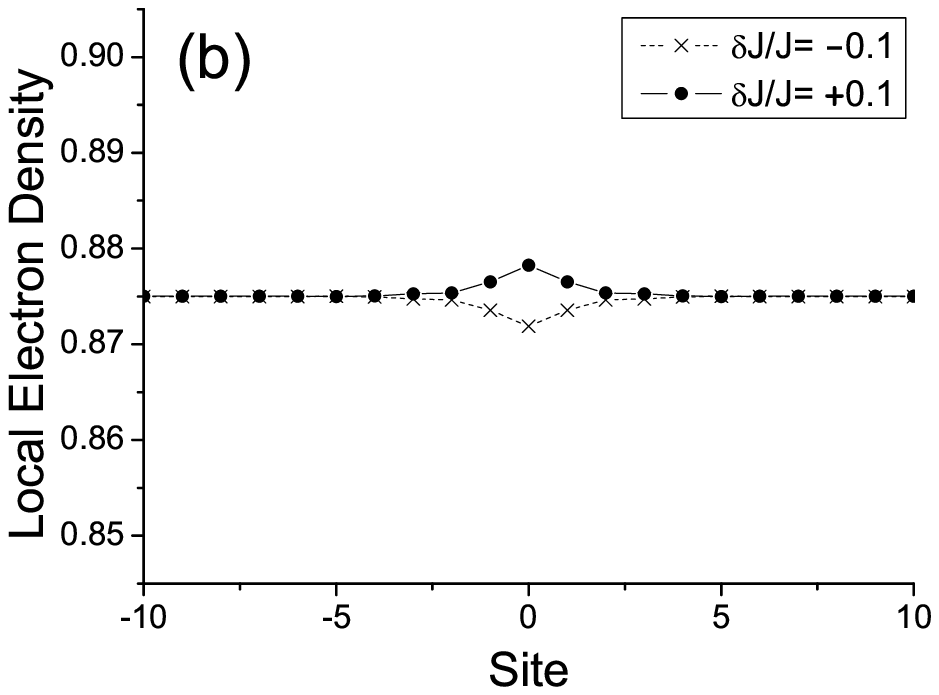}
 \end{center}
\caption{The local electron density along $x$-axis when an
impurity is located at 0: (a) a type-(ii) impurity ($t$-variation),
(b) a type-(iii) impurity ($J$-variation). }
\end{figure}




We thank C.-M. Ho for discussion.

\end{document}